# Fully Few-shot Class-incremental Audio Classification Using Expandable Dual-embedding Extractor

*Yongjie Si, Yanxiong Li\*, Jialong Li, Jiaxin Tan, Qianhua He*

School of Electronic and Information Engineering, South China University of Technology, Guangzhou, China
eeyongjiesi@mail.scut.edu.cn, eeyxli@scut.edu.cn

## Abstract

It's assumed that training data is sufficient in base session of few-shot class-incremental audio classification. However, it's difficult to collect abundant samples for model training in base session in some practical scenarios due to the data scarcity of some classes. This paper explores a new problem of fully few-shot class-incremental audio classification with few training samples in all sessions. Moreover, we propose a method using expandable dual-embedding extractor to solve it. The proposed model consists of an embedding extractor and an expandable classifier. The embedding extractor consists of a pretrained Audio Spectrogram Transformer (AST) and a finetuned AST. The expandable classifier consists of prototypes and each prototype represents a class. Experiments are conducted on three datasets (LS-100, NSynth-100 and FSC-89). Results show that our method exceeds seven baseline ones in average accuracy with statistical significance. Code is at: https://github.com/YongjieSi/EDE.

## 1. Introduction

Audio classification (AC) aims to recognize different classes of sounds in environment. It can be used for intelligent assisted driving [1], medical care [2], wildlife protection [3], ecological environment detection [4], and smart home [5].

Some works were done on audio classification in recent years. According to the ability to recognize new classes and the number of training samples required, the current works can be divided into four categories: many-shot AC [6]-[9], few-shot AC [10]-[12], class-incremental AC [13], [14], and Few-shot Class-incremental AC (FCAC) [15]-[18]. The many-shot AC method can recognize the pre-given classes only and requires many-shot training samples. The few-shot AC method can recognize new classes with few-shot training samples, but cannot remember old ones. The class-incremental AC method can recognize new classes with remembering old ones, but requires many-shot training samples in all sessions. The FCAC method requires abundant training samples in base session, but only few training samples in incremental session.

Although these efforts have contributed to the development of AC, they still leave something to be desired. For example, adequate training data is required for the FCAC methods in base session to learn an embedding extractor, in order to ensure better performance during incremental sessions. That is, the performance of these methods heavily depends on the availability of adequate training data.

In some practical scenarios (e.g., identification of criminal voices, classification of endangered animal calls), it is difficult to collect abundant samples for all classes. In this work, we aim to incrementally recognize new classes without forgetting old ones when training samples are few in all sessions, namely to address the problem of Fully FCAC (FFCAC). Main challenge in the task of FFCAC is that the model constructed in base session cannot be trained well due to the lack of training samples. Therefore, both the stability (ability to remember old classes) and plasticity (ability to continuously identify new classes) of the model are poor.

To address the above challenge, we design an Expandable Dual-embedding Extractor (EDE). The EDE is designed based on an Audio Spectrogram Transformer (AST) [19] which is an attention-based audio classification model. On the one hand, we finetune the AST on datasets of current task due to the distribution gap between the datasets used for pretraining and the datasets of current task. On the other hand, the finetuned AST lacks generalizability since it has been updated by training samples of current task. As a result, the model is prone to forgetting old classes in incremental sessions. Accordingly, we integrate the pretrained AST and the finetuned AST into a dual-embedding extractor to achieve embedding complementarity and improve the stability of the model. To further improve the plasticity of the model, the dual-embedding extractor is dynamically expanded to generate discriminative prototypes. Deep layers (layers closer to the output) of the network can extract task-specific embeddings [20]. Therefore, we expand the last layer of the EDE to learn diverse task-specific embeddings in each incremental session, while the remaining layers of the network are frozen. In each session, each prototype is obtained by computing the mean vector of embeddings extracted by the EDE. Classifier is composed of the prototypes of both new classes and old classes in each session.

Experiments are conducted on three public datasets, namely LS-100, NSynth-100, and FSC-89 [16]. Experimental results show that the proposed method exceeds previous methods in Average Accuracy (AA). The contributions of the work in this paper are summarized as follows.

1) We tackle a new problem of FFCAC, in which only few training samples are available for model training in all sessions. The setting of the FFCAC problem is in line with many practical scenarios.

2) We propose a FFCAC method using an EDE. The EDE

---

can be expandable and extract two kinds of embeddings, namely task-specific embedding and task-general embedding.

## 2. Method

### 2.1. Problem Definition

Like the definition of the FCAC [16], the FFCAC also includes $M$ sessions, namely one base session (session 0) and $M$-1 incremental sessions (sessions 1 to $M$-1). The training and testing datasets of different sessions are denoted by $\{\boldsymbol{D}_0^{tr}, \boldsymbol{D}_1^{tr}, ..., \boldsymbol{D}_m^{tr}, ..., \boldsymbol{D}_{M-1}^{tr}\}$ and $\{\boldsymbol{D}_0^{te}, \boldsymbol{D}_1^{te}, ..., \boldsymbol{D}_m^{te}, ..., \boldsymbol{D}_{M-1}^{te}\}$, respectively. $\boldsymbol{D}_m^{tr}$ and $\boldsymbol{D}_m^{te}$ have the same label set which is denoted by $\boldsymbol{L}_m$. The dataset in different sessions do not have the same type of classes, namely $\forall m, h$ and $m \neq h$, $\boldsymbol{L}_m \cap \boldsymbol{L}_h = \phi$. In the $m$th session, only $\boldsymbol{D}_m^{tr}$ can be used to train the model, and the trained model needs to be evaluated on the testing dataset of both current and all prior sessions, namely $\boldsymbol{D}_0^{te} \cup \boldsymbol{D}_1^{te} ... \cup \boldsymbol{D}_m^{te}$. Each one of all training datasets $\boldsymbol{D}_m^{tr}$ ($0 \leqslant m \leqslant (M\text{-}1)$) has few samples (small-scale dataset) and is constructed as a $N$-way $K$-shot training dataset. In the FCAC, however, $\boldsymbol{D}_0^{tr}$ has abundant samples (large-scale dataset) but $\boldsymbol{D}_m^{tr}$ ($1 \leqslant m \leqslant (M\text{-}1)$) has few samples (small-scale dataset).

### 2.2. Framework of the Method

Figure 1 shows the framework of our method which includes base and incremental sessions. There are three steps in the base session, including: finetune pretrained AST, construct EDE and generate base classifier. In each incremental session, the EDE is expanded in structure and the prototypes of the classifier are updated to recognize all seen audio classes.

In the base session, the AST finetuned in a supervised way on $\boldsymbol{D}_0^{tr}$ by minimizing the loss

$$L_{base}(\boldsymbol{x}, y; \psi) = -log \frac{\exp(\eta \cos(\psi(\boldsymbol{x}), \boldsymbol{w}_y))}{\sum_{n=1}^N \exp(\eta \cos(\psi(\boldsymbol{x}), \boldsymbol{w}_n))}, \quad (1)$$

where $\boldsymbol{x}$ and $y$ are input samples and the corresponding label, $\eta$ and $\boldsymbol{w}_n$ represent a scale factor and the classifier weight of class $n$, $\cos(\psi(\boldsymbol{x}), \boldsymbol{w}_n) = \frac{\psi(\boldsymbol{x}) \cdot \boldsymbol{w}_n}{\|\psi(\boldsymbol{x})\|_2 \cdot \|\boldsymbol{w}_n\|_2}$ stands for the cosine similarity between $\psi(\boldsymbol{x})$ and $\boldsymbol{w}_n$, $\|\cdot\|_2$ denotes 2-norm, $\psi(\boldsymbol{x})$ represents the embedding of $\boldsymbol{x}$. Then, the finetuned AST $\psi_0$ and the pretrained AST $\psi_{pre}$ are merged to construct the EDE $\{\psi_{pre}, \psi_0\}$. Next, the EDE is used to learn the embeddings from the Log Mel-spectra of the training samples. Finally, prototype for class $n$ is computed by

$$\boldsymbol{p}_n = \frac{1}{K} \sum_{i=1}^K \psi(\boldsymbol{x}), \quad (2)$$

where $\boldsymbol{p}_n$ and $K$ denote the prototype of class $n$ and the number of samples, respectively. The classifier consists of prototypes and each prototype represents one class. To further reduce the model's forgetting of old classes, we save the covariance matrix $\boldsymbol{\sigma}$ of embeddings for each of seen classes so far, and reconstruct the embeddings $\boldsymbol{e}'$ of each old class by

$$\boldsymbol{e}' = \boldsymbol{p} + \boldsymbol{\varepsilon}\boldsymbol{\sigma}^{-1}, \quad (3)$$

where $\boldsymbol{\varepsilon}$ is generated based on the standard normal distribution. $\boldsymbol{\sigma}^{-1}$ denotes the inverse matrix of $\boldsymbol{\sigma}$.

In the $m$th ($1 \leqslant m$) incremental session, the EDE is expanded and trained on $\boldsymbol{D}_m^{tr}$ and reconstructed embeddings. Then the expanded AST $\psi_m$ and the pretrained AST $\psi_{pre}$ are merged as the backbone of the EDE, namely $\{\psi_{pre}, \psi_m\}$. Next, the embeddings are extracted from the Log Mel-spectra of training samples by the updated EDE. New prototypes are generated to update the classifier together with old prototypes.

### 2.3. Expandable Dual-embedding Extractor

The motivation for designing EDE is that the embeddings extracted by the finetuned AST and the pre-trained AST are task-specific and task-general, respectively, and are therefore complementary. As shown in Figure 2, the EDE is a concatenation of the finetuned AST and the pre-trained AST. The log Mel-spectrum of each audio sample is split into patches and added with a trainable positional embedding before being fed to the transformer encoder. The transformer encoder consists of several blocks with multi-head attention. There are two differences between the AST used here and the original AST. First, we apply mean pooling over all patch representation as the embedding of the audio clip instead of appending a token at the beginning of the sequence to summarize all patch representations. Second, we replace the linear classifier with a cosine head, by which the maximum logits are bounded in the range of [-1, 1] and are expected to be more discriminative.

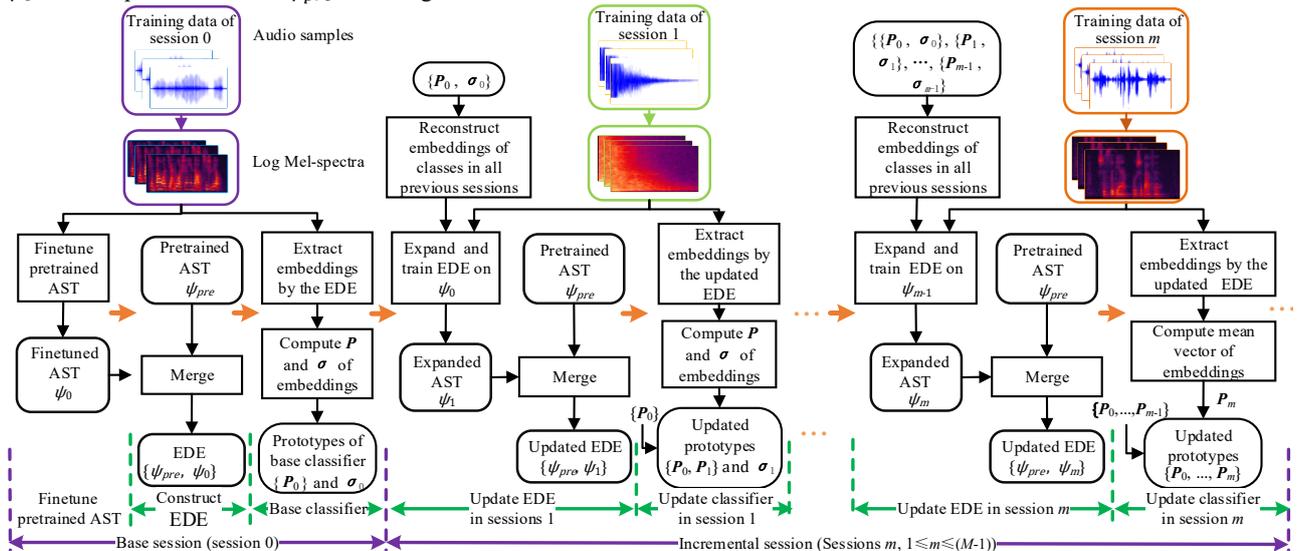

Figure 1: *The framework of our method which consists of M sessions. AST: Audio Spectrogram Transformer. EDE: Expandable Dual-embedding Extractor. ψ: parameters of the EDE. **P**: mean vector of embeddings. **σ**: covariance of embeddings.*

The AST which is fine-tuned on a few samples of the current task lacks the ability to learn a discriminative embedding to effectively represent each class of the current task. Therefore, it is desirable to improve the representational ability of the embedding extractor. It is shown that the shallow layers and the deep layers of a deep neural network are able to extract task general embeddings and task specific embeddings, respectively, from the input samples [20]. Accordingly, we decompose the transformer encoder into generalizable (shallow) layers $\psi_g$ and specialized (deep) layers $\psi_s$, i.e., $\psi(x) = \psi_s(\psi_g(x))$. Specifically, in the $m$th (1≤$m$) incremental session, the AST finetuned in the previous session $\psi_{old} = \{\psi_{s_0}(\psi_{g_0}(x)), \cdots, \psi_{s_{m-1}}(\psi_{g_0}(x))\}$ is frozen to retain the knowledge of old classes, the last layer of the transformer encoder is extended by a block $\psi_{s_m}$, which is initialized with $\psi_{s_{m-1}}$ and trained on $D_m^{tr}$ and the reconstructed embeddings by minimizing the loss

$$L_{inc}(x, y; \psi) = L_{base}(x, y; \psi) - \lambda \log \frac{\exp(\eta \cos(\psi(x), w_n^{aux}))}{\sum_{n=1}^{N} \exp(\eta \cos(\psi(x), w_n^{aux}))} \quad (4)$$

where $\psi(x) = \{\psi_{old}(x), \psi_m(x)\}$, $w_n^{aux}$ is the weight of an auxiliary classifier, $\lambda$ is used to control the effect of the auxiliary classifier.

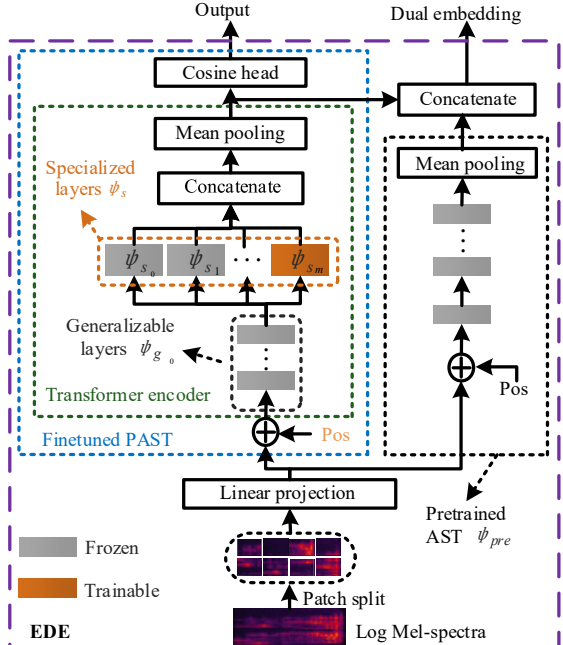

Figure 2: *The architecture of EDE, ψ: parameters of the EDE.*

## 3. Experiments

### 3.1. Experimental Datasets

Experiments are carried out on three datasets, including the LS-100, NSynth-100 and FSC-89. They are publicly available and are widely used in prior work. Their details are given at the three websites[1]. We randomly select audio samples of 25 classes from each one dataset. The selected samples are split into $M$ parts without overlaps of classes, namely $D_m$ (0 ≤ $m$ ≤ ($M$-1)). $D_m$ is composed of training dataset $D_m^{tr}$ and testing dataset $D_m^{te}$. Table 1 presents the detailed information of the datasets of LS-100, NSynth-100 and FSC-89.

---

[1]https://www.modelscope.cn/datasets/pp199124903/LS-100/summary
https://www.modelscope.cn/datasets/pp199124903/FSC-89/summary
https://www.modelscope.cn/datasets/pp199124903/NSynth-100/summary

Table 1: *Detailed information of LS-100/ NSynth -100/FSC-89.*

| Parameters | $D_m$ | |
|---|---|---|
| | $D_m^{tr}$ | $D_m^{te}$ |
| #Classes | 100/100/89 | 100/100/89 |
| Length (hours) | 9.72/6.67/13.11 | 3.33/1.52/3.28 |
| #Samples/Class | 500/200/800 | 100/100/200 |

### 3.2. Experimental Setup

AA is an assessment of the overall performance of different methods and is defined by

$$AA = \frac{1}{M} \sum_{m=0}^{M-1} A_m \quad (5)$$

where $A_m$ stands for the accuracy in session $m$.

The AST is pretrained on both ImageNet and AudioSet. In each training step, $N \cdot K$ samples are randomly chosen from $D_m^{tr}$ to train the model in all sessions. All testing datasets of seen classes, namely $D_0^{te} \cup \cdots \cup D_m^{te}$, are used to evaluate the model's performance. The process is repeated 100 times to obtain a result that consists of a mean value and a standard deviation of the AA. Dimensions Log of Mel spectrum and prototype are set to 128 and 768 respectively. In each session, the model is trained for 100 epochs. The learning rate starts at 0.001 and decays with cosine annealing. The value of ($N$, $K$) is set to (5, 5) without loss of generality, and $\lambda$ is set to 1.

### 3.3. Ablation Experiments

In this subsection, to verify the effectiveness of the main components of the proposed method, we perform ablation analyses on the samples of the selected 25 classes in each of the LS-100, NSynth-100 and FSC-89. We discuss the impacts of the pretrained AST (P-AST), the finetuned AST (F-AST) and the expanded F-AST. Table 2 shows the results obtained by our method with different combinations of P-AST, F-AST and extended F-AST. Our method achieves the highest AA scores on three experimental datasets when using P-AST and extended F-AST. Namely, the proposed method achieves the highest AA scores of 61.85%, 65.40% and 37.36% for all classes.

Table 2: *Results obtained by our method on three datasets with different combinations of P-AST, F-AST and expanded F-AST.*

| P-AST | F-AST | expanded F-AST | AA (%) | | |
|---|---|---|---|---|---|
| | | | LS-100 | NSynth-100 | FSC-89 |
| √ | × | × | 60.09 | 54.58 | 34.73 |
| × | √ | × | 60.31 | 64.00 | 36.94 |
| √ | √ | × | 61.24 | 64.82 | 37.20 |
| √ | × | √ | 61.85 | 65.40 | 37.36 |

### 3.4. Comparison of Different Methods

In this subsection, our method is compared with seven baseline methods. These methods are labelled as Finetune, iCaRL [24], DER [25], PODNET [26], CEC [27], FACT [28] and PAN [16]. The Finetune method has a tendency to overfit the new classes and forget the old ones after fine-tuning with training samples of the new classes. The iCaRL method uses data retention and knowledge distillation to train the embedding extractor and classifier to mitigate the forgetting of old classes. The DER method creates a new backbone for each updating sessions. The PODNET method proposes a spatially based distillation loss and a multiple proxy vector representation to balance the adaptivity and generalizability of the model. The CEC method designs a graph model to propagate the contextual information between classifiers. The FACT method builds a prototype-based classifier and reserves the embedding space for new classes in incremental sessions. The PAN method designs self-attention modified prototypes to dynamically expand the classifier.

All the basic methods are carried out under the same conditions, and the values of the AA obtained by the different methods on the experimental datasets are presented in Tables 3 to 5. Our method achieves AA scores of 61.85%, 65.40% and 37.36% on the audio samples from LS-100, NSynth-100 and FSC-89, respectively. These three AA scores are higher than the counterparts obtained by all baseline methods. The advantage of our method is mainly due to the use of EDE, which is extensible and fuses dual embedding.

In addition, we conduct the significance test for the comparison of different methods. Specifically, we analyze statistical significance using Friedman test (null-hypothesis test) [29] with Nemenyi test (post-hoc test) [30] which has been widely adopted for significance test. The significance test is performed on the average rank of the 100 test results per experimental dataset. Figure 3 (a) shows the number of times each method achieves each rank. Figure 3 (b) shows the results of the significance test for a confidence level $\alpha$ of 0.05. The digits from 1 to 8 represent mean rank for different methods. The method corresponding to small numerical values is superior to the method corresponding to large numerical values. The methods spanned by the thick black-crossbars do not have sufficient evidence of statistically significant differences. It can be concluded from Figure 3 that our method has advantage over baseline methods in accuracy, and this accuracy advantage is statistically significant.

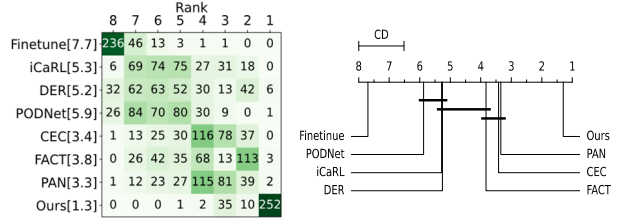

*(a) Histogram of method ranks   (b) Critical difference (CD) diagram*
Figure 3: *Results of significance test. (a) Visualization of the number of times each method achieves each rank. (b) Visualization of significance differences between various methods.*

Table 3: *Results obtained by different methods on LS-100.*

| Methods | Accuracy in various sessions (%) | | | | | AA (%) |
|---|---|---|---|---|---|---|
| | 0 | 1 | 2 | 3 | 4 | |
| Finetune | 73.56±8.87 | 29.71±4.92 | 13.82±3.52 | 10.49±3.10 | 10.07±3.37 | 27.53±3.07 |
| iCaRL | 73.15±8.78 | 34.40±7.86 | 21.75±5.07 | 18.30±3.65 | 17.77±3.19 | 33.07±4.32 |
| DER | 73.36±8.22 | 37.36±7.20 | 22.95±7.83 | 19.95±7.01 | 18.44±6.51 | 34.41±6.55 |
| PODNET | 73.09±7.96 | 38.54±5.42 | 27.61±4.48 | 24.03±3.53 | 24.56±2.85 | 37.57±3.39 |
| CEC | 85.40±6.40 | 52.17±6.18 | 39.23±4.82 | 32.95±3.75 | 29.78±3.46 | 47.91±4.10 |
| FACT | 88.41±5.41 | 55.59±4.99 | 41.34±4.47 | 33.83±3.95 | 29.65±3.92 | 49.76±4.02 |
| PAN | 85.70±6.19 | 52.20±6.16 | 39.17±4.84 | 32.95±3.74 | 29.88±3.45 | 47.98±4.23 |
| Ours | 91.90±3.73 | 70.23±3.30 | 54.21±2.09 | 46.97±2.53 | 45.94±2.25 | **61.85±2.75** |

Table 4: *Results obtained by different methods on NSynth-100.*

| Methods | Accuracy in various sessions (%) | | | | | AA (%) |
|---|---|---|---|---|---|---|
| | 0 | 1 | 2 | 3 | 4 | |
| Finetune | 71.88±5.07 | 52.60±5.57 | 34.74±3.89 | 27.11±3.52 | 24.18±3.08 | 42.10±2.31 |
| iCaRL | 71.70±5.12 | 53.51±4.71 | 53.66±4.14 | 49.07±3.46 | 49.48±3.33 | 55.48±3.70 |
| DER | 74.40±5.53 | 61.42±6.14 | 60.06±6.07 | 53.73±7.47 | 44.42±6.50 | 58.81±4.92 |
| PODNET | 71.87±6.83 | 44.89±6.01 | 43.58±5.26 | 42.93±5.08 | 40.97±5.96 | 48.85±4.50 |
| CEC | 76.13±6.03 | 58.15±5.51 | 53.80±4.52 | 48.26±4.69 | 44.34±5.11 | 56.14±4.14 |
| FACT | 74.97±4.92 | 51.94±4.91 | 51.43±4.50 | 46.54±4.04 | 43.45±3.65 | 53.27±4.51 |
| PAN | 76.71±5.85 | 58.38±5.47 | 53.92±4.43 | 48.44±4.64 | 44.48±5.03 | 56.39±4.20 |
| Ours | 76.16±5.49 | 70.18±4.28 | 63.46±3.23 | 59.16±3.61 | 58.02±3.05 | **65.40±3.80** |

Table 5: *Results obtained by different methods on FSC-89.*

| Methods | Accuracy in various sessions (%) | | | | | AA (%) |
|---|---|---|---|---|---|---|
| | 0 | 1 | 2 | 3 | 4 | |
| Finetune | 30.66±3.18 | 21.56±2.06 | 12.55±2.23 | 11.45±1.49 | 8.49±1.20 | 16.94±1.19 |
| iCaRL | 31.27±3.15 | 20.79±2.34 | 16.86±2.43 | 15.44±1.78 | 12.60±1.32 | 19.39±3.70 |
| DER | 31.12±3.02 | 21.39±2.44 | 15.63±2.14 | 12.93±1.52 | 9.14±1.08 | 18.04±1.10 |
| PODNET | 35.64±4.35 | 20.46±2.40 | 14.50±1.90 | 13.87±1.72 | 10.89±1.27 | 19.07±1.72 |
| CEC | 41.23±4.57 | 23.63±2.91 | 17.99±2.33 | 15.22±1.97 | 12.21±1.63 | 22.06±2.24 |
| FACT | 46.35±3.66 | 26.18±2.39 | 22.87±2.18 | 20.13±1.78 | 16.53±1.35 | 26.41±1.86 |
| PAN | 41.48±4.36 | 23.72±2.87 | 18.08±2.24 | 15.27±1.93 | 12.25±1.60 | 22.16±2.10 |
| Ours | 53.25±3.84 | 37.65±2.39 | 35.59±2.10 | 32.69±1.87 | 27.60±1.66 | **37.36±1.84** |

## 4. Conclusions

In this paper, we discussed a new problem of FFCAC, and tried to solve this problem by designing an expandable dual-embedding extractor. Based on the description of our method and experimental evaluations, we can draw two conclusions. First, our method outperforms state-of-the-art methods in terms of AA under the same experimental conditions. Second, we designed an EDE to improve the performance of the model. Although our method possesses advantages over all baseline ones in AA, it still needs to be improved. For instance, the EDE is based on a pre-trained AST which requires a relatively large amount of memory, and more memory will be required as it expands. In future work, we will consider reducing memory overhead of our method.

## 5. Acknowledgements

This work was partly supported by the national natural science foundation of China (62371195, 62111530145, 61771200), international scientific research collaboration project of Guangdong (2023A0505050116), Guangdong basic and applied basic research foundation (2022A1515011687), and Guangdong provincial key laboratory of human digital twin (2022B1212010004).

## 6. References

[1] T. Li, K. Kaewtip, J. Feng and L. Lin, "IVAS: Facilitating safe and comfortable driving with intelligent vehicle audio systems," in *Proc. of IEEE International Conference on Big Data (Big Data)*, 2018, pp. 5381-5382.

[2] K. Vatanparvar, V. Nathan, E. Nemati, M. M. Rahman and J. Kuang, "A generative model for speech segmentation and obfuscation for remote health monitoring," in *Proc. of IEEE International Conference on Wearable and Implantable Body Sensor Networks (BSN)*, 2019, pp. 1-4.

[3] A. Terenzi, N. Ortolani, I. Nolasco, E. Benetos and S. Cecchi, "Comparison of feature extraction methods for sound-based classification of honey bee activity," *IEEE/ACM TASLP*, vol. 30, pp. 112-122, 2022.

[4] X. Zhou, K. Hu and Z. Guan, "Environmental sound classification of western black-crowned gibbon habitat based on spectral subtraction and VGG16," in *Proc. of IEEE Advanced Information Management, Communicates, Electronic and Automation Control Conference (IMCEC)*, 2022, pp. 578-582.

[5] S. Nandhini, H. Rubla, P. Petchiammal, et al., "IoT based smart home with voice controlled appliances using raspberry Pi," in *Proc. of ICOEI*, 2022, pp. 624-629.

[6] Y. Wang, G. Zhao, K. Xiong, G. Shi and Y. Zhang, "Multi-scale and single-scale fully convolutional networks for sound event detection," *Neurocomputing*, vol. 421, pp. 51-65, 2021.

[7] J. Yan, Y. Song, L.-R. Dai and I. McLoughlin, "Task-aware mean teacher method for large scale weakly labeled semi-supervised sound event detection," in *Proc. of IEEE ICASSP*, 2020, pp. 326-330.

[8] H.P. Huang, K.C. Puvvada, M. Sun and C. Wang, "Unsupervised and semi-supervised few-shot acoustic event classification," in *Proc. of IEEE ICASSP*, 2021, pp. 331-335.

[9] J. Wu, F. Yang and W. Hu, "Unsupervised anomalous sound detection for industrial monitoring based on ArcFace classifier and Gaussian mixture model," *Applied Acoustics*, vol. 203, pp. 1-11, 2023.

[10] Y. Wang and D.V. Anderson, "Hybrid attention-based prototypical networks for few-shot sound classification," in *Proc. of IEEE ICASSP*, 2022, pp. 651-655.

[11] D. Yang, H. Wang, Y. Zou, Z. Ye and W. Wang, "A mutual learning framework for few-shot sound event detection," in *Proc. of IEEE ICASSP*, 2022, pp. 811-815.

[12] T. Zhang, L. Yang, X. Gu and Y. Wang, "A task-specific meta-learning framework for few-shot sound event detection," in *Proc. of IEEE MMSP*, 2022, pp. 1-6.

[13] E. Koh, F. Saki, Y. Guo, C. Hung and E. Visser "Incremental learning algorithm for sound event detection," in *Proc. of IEEE ICME*, 2020, pp. 1-6.

[14] D. Ma, C.I. Tangy and C. Mascolo, "Improving feature generalizability with multitask learning in class incremental learning," in *Proc. of IEEE ICASSP*, 2022, pp. 4173-4177.

[15] Y. Wang, N. J. Bryan, M. Cartwright, J. Pablo Bello and J. Salamon, "Few-shot continual learning for audio classification," in *Proc. of IEEE ICASSP*, 2021, pp. 321-325.

[16] Y. Li, W. Cao, W. Xie, J. Li and E. Benetos, "Few-shot class-incremental audio classification using dynamically expanded classifier with self-attention modified prototypes," *IEEE TMM*, vol. 26, pp. 1346-1360, 2024.

[17] W. Xie, Y. Li, Q. He, W. Cao and T. Virtanen, "Few-shot class-incremental audio classification using adaptively-refined prototypes," in *Proc. of INTERSPEECH*, 2023, pp. 301-305.

[18] Y. Li, W. Cao, J. Li, W. Xie and Q. He, "Few-shot class-incremental audio classification using stochastic classifier," in *Proc. of INTERSPEECH*, 2023, pp. 4174-4178.

[19] Y. Gong, Y.-A. Chung and J. Glass, "AST: Audio Spectrogram Transformer," in *Proc. of Interspeech*, 2021, pp. 571-575.

[20] H. Maennel, I.M. Alabdulmohsin, I.O. Tolstikhin, R. Baldock, O. Bousquet, S. Gelly and D. Keysers, "What do neural networks learn when trained with random labels?" in *Proc. of NeurIPS*, 2020, vol. 33, pp. 19693-19704.

[21] J. Snell, K. Swersky and R. Zemel, "Prototypical networks for few-shot learning," in *Proc. of Advances in Neural Information Processing Systems*, 2017, pp. 4077-4087.

[22] V. Panayotov, G. Chen, D. Povey and S. Khudanpur, "Librispeech: An ASR corpus based on public domain audio books," in *Proc. of IEEE ICASSP*, 2015, pp. 5206-5210.

[23] J. Engel, C. Resnick, A. Roberts, S. Dieleman, M. Norouzi, D. Eck and K. Simonyan, "Neural audio synthesis of musical notes with WaveNet autoencoders," in *Proc. of ICML*, 2017, vol. 70, pp. 1068-1077.

[24] S.-A. Rebuffi, A. Kolesnikov, G. Sperl and C.H. Lampert, "iCaRL: Incremental classifier and representation learning," in *Proc. of IEEE CVPR*, 2017, pp. 5533-5542.

[25] Shipeng Yan, Jiangwei Xie and Xuming He, "Der: Dynamically expandable representation for class incremental learning," in *Proc. of CVPR*, 2021, pp. 3014-3023.

[26] A. Douillard, M. Cord, C. Ollion, T. Robert and E. Valle, "Podnet: Pooled outputs distillation for small-tasks incremental learning," in *Proc. of ECCV*, 2020, pp. 86-102.

[27] C. Zhang, N. Song, G. Lin, Y. Zheng, P. Pan and Y. Xu, "Few-shot incremental learning with continually evolved classifiers," in *Proc. of IEEE/CVF CVPR*, 2021, pp. 12450-1245.

[28] D.-W. Zhou, F.-Y. Wang, H.-J. Ye, L. Ma, S. Pu and D.-C. Zhan, "Forward compatible few-shot class-incremental learning," in *Proc. of IEEE/CVF CVPR*, 2022, pp. 9046-9056.

[29] J. Demšar, "Statistical Comparisons of Classifiers over MultipleData Sets," *J. Machine Learning Research*, vol. 7, pp. 1-30, 2006.

[30] S. García and F. Herrera, "An extension on 'statistical comparisons of classifiers over multiple data sets' for all pairwise comparisons," *J. Machine Learning Research*, vol. 9, pp. 2677-2694, 2008.